\documentclass[prd,aps,a4paper,superscriptaddress,onecolumn,nofootinbib]{revtex4}
\usepackage{graphicx}
\usepackage{color}
\usepackage{dcolumn}
\usepackage{bm}
\usepackage{slashed}
\usepackage{amsmath}
\usepackage{latexsym}
\usepackage{amssymb}
\usepackage{mathrsfs}
\usepackage{amsfonts}
\usepackage{url}
\usepackage{graphicx}

\newcommand{\nnm}{\nonumber}

\newcommand{\be}{\begin{equation}}
\newcommand{\ee}{\end{equation}}

\newcommand{\mr}{\mathrm}

\newcommand{\mc}{\mathcal}

\newcommand{\bs}{\boldsymbol}

\allowdisplaybreaks
\begin{document}
	
\title{Effective metric for bound state in an effective-one-body \\ theory based on the third-post-Minkowskian approximation}

\author{Hanjun Zou}
\email[Hanjun Zou:~]{zouhanjun24@mails.ucas.ac.cn}
\affiliation{School of Fundamental Physics and Mathematical Sciences, Hangzhou Institute for Advanced Study, University of Chinese Academy of Sciences, 
Hangzhou 310024, China}
\author{Sheng Long}
\email[Sheng Long:~]{shenglong@mails.ucas.ac.cn}
\affiliation{School of Fundamental Physics and Mathematical Sciences, Hangzhou Institute for Advanced Study, University of Chinese Academy of Sciences, 
Hangzhou 310024, China}
\author{Xiaokai He}
\email[Xiaokai He:~]{sjyhexiaokai@hnfnu.edu.cn}
\affiliation{School of Mathematics and Statistics, Hunan First Normal University, Changsha 410205, China}
\author{Zhoujian Cao\footnote{corresponding author}}
\email[Zhoujian Cao:~]{zjcao@bnu.edu.cn}
\affiliation{School of Fundamental Physics and Mathematical Sciences, Hangzhou Institute for Advanced Study, University of Chinese Academy of Sciences, 
Hangzhou 310024, China}
\affiliation{School of Physics and Astronomy, Beijing Normal University, Beijing 100875, China}

\begin{abstract}
The effective-one-body (EOB) framework, originally formulated within the post-Newtonian (PN) expansion, is central to modeling and interpreting gravitational-wave signals. Recent developments have incorporated  the post-Minkowskian (PM) expansion into EOB theory. In this work, we focus on the bound state dynamics in the PM approximation up to the third order and construct the core ingredients of EOB: the effective metric. We examine two correspondence strategies, one based on the radial action variable and the other on the precession angle. We demonstrate that the radial action variable provides a consistent correspondence, which is further verified by the correspondence based on the precession angle. Building on these results, we adopt an isotropic gauge with a Schwarzschild-like parametrization to fix the remaining freedom. Within this parametrization, the effective metric coefficients are determined at 3PM order. Our results provide a consistent effective metric for the bound state dynamics.
\end{abstract}

\maketitle

\section{Introduction}

The advent of gravitational wave (GW) astronomy~\cite{abbott2016observation,LIGO,abbott2017gw170817,abbott2019gwtc}, initiated by the first direct detection by LIGO~\cite{LIGO}, has opened an unprecedented window into the cosmos, allowing for novel investigations into astrophysics and fundamental physics~\cite{cai2017gravitational}. The success of these GW detections relies heavily on the matched filtering technique, which requires highly accurate theoretical models of gravitational waveforms emitted by sources like coalescing compact binaries~\cite{taracchini2014effective}.

\par Modeling the complete inspiral-merger-ringdown sequence of a binary system presents a significant theoretical challenge, as analytical and numerical methods are best suited for different phases of the evolution. The post-Newtonian (PN) approximation, a weak-field and slow-motion ($v/c \ll 1$) expansion, effectively describes the early inspiral phase when the bodies are well-separated. Conversely, numerical relativity (NR) simulations can accurately model the highly non-linear, strong-field dynamics of the late inspiral and merger.

\par The Effective-One-Body (EOB) formalism within the Post-Minkowskian (PM) framework was developed to bridge this gap \cite{damour1PM2016,damour2PM2019,he2021energy,Sun2021gka,Jing2021ahx,Jing2022vks,Jing2023okh,Long2023vph,Long2024axi,Deng2024ayh,Jing2025utt}. Its core idea is to map the complicated two-body problem onto the simpler dynamics of a single effective particle moving in a deformed background spacetime. And the PM expansion further ensures that the resulting predictions remain accurate in the relativistic, fast-motion regime.

\par More specifically, to determine the deformed background spacetime metric we need to build the correspondence between the real two-body system and the effective one-body system which requires specifying appropriate dictionaries. A key dictionary is the energy map between the real relativistic energy $\mathcal{E}$ (in the center of mass frame) and the effective one $\mathcal{E}_0$, which was developed by Damour (2016) within the PM approximation \cite{damour1PM2016}:
\be\label{eq::EnergMap}
\frac{\mathcal{E}_0}{m_0 c^2}
= \frac{\mathcal{E}^2 - m_1^2 c^4 - m_2^2 c^4}{2 m_1 m_2 c^4}\,.
\ee
Beyond the energy map, there are two standard ways to construct a second correspondence for the bound state dynamics. The first equates the radial action variable $I_\text{R}$ computed from the Hamiltonian of the real two-body system to the effective radial action $I^{\text{eff}}_\text{R}$ of a test particle moving in a deformed Schwarzschild spacetime. The second equates the precession angle, by matching the precession angle computed in the real system to the corresponding effective angle obtained in the deformed Schwarzschild spacetime. Fortunately, profiting from recent advances in the area of scattering amplitudes, the conservative real Hamiltonian for a two-body system has been recently derived at 2PM order  and 3PM order \cite{bern2019scattering,bern2019long}. This allows us to use it to calculate the radial action variable and precession angle of a real two-body system. The only thing left is to find the effective radial action variable $I^{\text{eff}}_\text{R}$ and effective precession angle $\psi^{\rm{eff}}$.

\par It is important to emphasize that the most frequent sources of GWs for LIGO/ Virgo experiments are bound/inspiraling binaries, instead of unbound/scattering ones. Many previous studies have relied on the correspondence between the scattering functions $\chi^{\text{real}}_{\text{PM}} $ and $ \chi^{\text{eff}}_{\text{PM}}$,
often incorporating Finsler-type terms to account for higher-order corrections \cite{damour2PM2019,antonelli2019energetics,Jing2023okh}. Recently developed methods have enabled the mapping of scattering processes to bound states \cite{Kalin2019rwq,Kalin2019inp,Jing2025ihz}. However, our work diverges from this approach by focusing exclusively on bound states directly. More precisely, we utilize the radial action and the precession angle of closed orbit, which is distinctive to bound systems, to carry out our calculations. Furthermore, instead of employing the Finsler-type corrections, we expand the coefficients of the deformed Schwarzschild metric by order $G$ to complete the correspondence of higher-order terms. We also note the recently proposed eXact One-Body (XOB) approach~\cite{Zeng2024XOB}, which seeks a one-body description whose conservative sector requires no PN approximation as input and is intended to cover the inspiral, merger, and ringdown stages of black-hole binary evolution. In contrast, our perturbative 3PM construction describes conservative bound dynamics by matching the radial action and precession angle. The two approaches are conceptually related but methodologically complementary.

\par The rest of the paper is organized as follows. In Sec.~\ref{sec::RadialAction}, we show that the radial action variable $I_\text{R}$ from the Hamiltonian of the real two-body system at 3PM accuracy should be sought from higher-order corrections, and the same thing happens to the effective radial interaction variable $I^{\text{eff}}_\text{R}$ of a test particle moving in a deformed Schwarzschild spacetime (effective background spacetime). This result is consistent with the result obtained from the precession angle.
In Sec.~\ref{sec::PrecessionAngle}, we calculate the precession angle of a real two-body system. After imposing a Schwarzschild-like parametrization we get the effective metric at 3PM order by comparing the precession angle of a real two-body system and the EOB system.
Finally, a summary and discussion are given in Sec.~\ref{sec::Summary}.

\par We use the unit with c = 1 throughout the paper, which is suitable for calculations within the PM framework.

\section{Energy map and effective metric based on action variable}
\label{sec::RadialAction}

In this section, we try to calculate the radial action variable both for a real two-body system and the EOB system at 3PM order, respectively. We point out that the results of 3PM accuracy should be sought from higher-order corrections in the radial action.

\subsection{Radial action variable of real two-body system at 3PM order}

In 2019, Bern et al.~\cite{bern2019scattering} derived an explicit form of the conservative Hamiltonian for a nonspinning, massive two-body system at third post-Minkowskian (3PM) order:
\be
H(\bs r,\bs p)=H_0(\bs p^2)
+\sum_{n=1}^3\frac{G^n}{r^n}c_n(\bs p^2)
+\mc O(G^4),
\label{3PMH}
\ee
where the free Hamiltonian part reads
\be
H_0(\bs p^2)=\sqrt{m_1^2+\bs p^2}+\sqrt{m_2^2+\bs p^2}.
\label{H0}
\ee
Here, $\bs r$ denotes the relative radial separation and $\bs p$ is its conjugate momentum, respectively. The coefficient functions $c_1$, $c_2$ and $c_3$ are provided explicitly in Eq.~(10) of~\cite{bern2019scattering}. With spins neglected, the motion is confined to a fixed orbital plane. Introducing polar coordinates $(r,\phi)$ within this plane, with conjugate variables $(p_r,p_\phi\equiv L)$ satisfying the standard relation
\be\label{eq::psq}
\bs p^2=p_r^2+\frac{L^2}{r^2}.
\ee
We identify the conserved angular $L=p_\phi$ by axial symmetry and denote with $\mathcal{E} = H(\bs r,\bs p)=H(r,p_r,L)$ the total conserved energy of the binary system. Then we can use the Hamilton-Jacobi approach to solve the 3PM dynamics. In polar coordinates, the Hamilton-Jacobi equation becomes
\be\label{eq::E}
\mc{E}=H_0(p_r^2+\frac{L^2}{r^2})+\sum_{n=1}^3\frac{G^n}{r^n}c_n(p_r^2+\frac{L^2}{r^2})
+\mc O(G^4).
\ee
We can expand $p_r^2$ as
\be\label{eq::prEp}
p_r^2 = \frac{P_0r^2 - L^2}{r^2} + P_1(\frac{G}{r}) + P_2(\frac{G}{r})^2 + P_3(\frac{G}{r})^3 + \mc O(G^4).
\ee
Substituting Eq.~\eqref{eq::prEp} into Eq.~\eqref{eq::E}, and using the free Hamiltonian in Eq.~\eqref{H0} together with Eq.~\eqref{eq::psq}, we expand the resulting Hamilton--Jacobi equation in powers of \(G\). Equating the coefficients at each order from \(G^0\) to \(G^3\) yields a sequence of algebraic equations that determines \(P_0\), \(P_1\), \(P_2\), and \(P_3\) successively. Using the explicit coefficients \(c_n(\bs p^2)\) given in Eq.~(10) of Ref.~\cite{bern2019scattering}, we obtain
\begin{widetext}
\begin{subequations}\label{eq::PValue}
\begin{align}
P_0 =& \frac{\mc{E}^4 + (m_1^2 - m_2^2)^2 - 2\mc{E}^2(m_1^2 + m_2^2)}{4\mc{E}^2}, \\
P_1 =& \frac{\mc{E}^4 - 2\mc{E}^2(m_1^2 + m_2^2) + (m_1^4 + m_2^4)}{\mc{E}},\\
P_2 =& \frac{3(m_1 + m_2)}{8\mc{E}} \left[ 5\mc{E}^4 - 5m_1^4 + 5m_2^4+ 6m_1^2 m_2^2 - 10\mc{E}^2 \left( m_1^2 + m_2^2 \right) \right],\\
P_3 =& \frac{1}{12 \mc{E} (m_1 + m_2 + \mc{E})(-(m_1 - m_2)^2 + \mc{E}^2 )} \Bigg\{ -28 m_1^9 +
  m_1^8 ( 28 m_2 + 17 \mc{E} ) +
  m_1 (m_2 - \mc{E})^2 (m_2 + \mc{E})^4 (28 m_2^2 - \mc{E}^2) -
  \nnm\\
  &(m_2 - \mc{E})^3 (m_2 + \mc{E})^4 (28 m_2^2 - 45 m_2 \mc{E} + 44 \mc{E}^2) +
  m_1^7 ( -256 m_2^2 + 56 m_2 \mc{E} + 85 \mc{E}^2 ) +
  \nnm\\
  & m_1^6 ( 256 m_2^3 - 258 m_2^2 \mc{E} - 29 m_2 \mc{E}^2 - 95 \mc{E}^3 ) +
  m_1^5 \mc{E} ( 568 m_2^3 + 547 m_2^2 \mc{E} - 114 m_2 \mc{E}^2 - 87 \mc{E}^3 ) -
  \nnm\\
  & m_1^2 (m_2 - \mc{E}) (m_2 + \mc{E})^2 ( 256 m_2^4 + 2 m_2^3 \mc{E} -
     293 m_2^2 \mc{E}^2 + 178 m_2 \mc{E}^3 - 149 \mc{E}^4 ) +
  \nnm\\
  & m_1^3 (m_2 + \mc{E})^2 ( 256 m_2^4 + 56 m_2^3 \mc{E} - 347 m_2^2 \mc{E}^2 -
     2 m_2 \mc{E}^3 + 31 \mc{E}^4 ) +
  \nnm\\
  & m_1^4 \mc{E} ( -478 m_2^4 + 21 m_2^3 \mc{E} + 373 m_2^2 \mc{E}^2 -
     27 m_2 \mc{E}^3 + 183 \mc{E}^4 ) +
  \nnm\\
  & \frac{1}{m_1 + m_2 - \mc{E}}48 \sqrt{ (m_1 - m_2 - \mc{E})(m_1 + m_2 - \mc{E})(m_1 - m_2 + \mc{E})(m_1 +
      m_2 + \mc{E})}
   \nnm\\
   & \left[ m_1^8 + (m_2^2 - \mc{E}^2)^4 -
     4 m_1^6 (2 m_2^2 + \mc{E}^2) +
     6 m_1^4 ( -5 m_2^4 + 2 m_2^2 \mc{E}^2 + \mc{E}^4 ) -
     4 m_1^2 ( 2 m_2^6 - 3 m_2^4 \mc{E}^2 + \mc{E}^6 ) \right]
   \nnm\\
   & \sinh^{-1}\left( \frac{1}{2} \sqrt{\frac{ - (m_1 + m_2)^2 + \mc{E}^2}{m_1 m_2}} \right) \Bigg\}.
\end{align}
\end{subequations}
\end{widetext}
According to the Hamilton-Jacobi theory, the radial effective potential $\mc{R}(r, \mc{E}, L )$ at the 3PM order reads
\be
\mc{R}(r, \mc{E}, L ) = \frac{P_0r^2 - L^2}{r^2} + P_1(\frac{G}{r}) + P_2(\frac{G}{r})^2 + P_3(\frac{G}{r})^3.
\ee
From the definition of the radial action variable, we have
\be\label{eq::IR}
I_\text{R} \equiv \frac{2}{2\pi} \int_{r_{\min}}^{r_{\max}} \mr{d}r \sqrt{\mc{R}(r, \mc{E}, L)},
\ee
where $r_{\min}$ and $r_{\max}$ are the turning points of the effective potential $\mc{R}(r, \mc{E}, L )$. In general, it is hard to get the exact expression of Eqs.~\eqref{eq::IR} by direct integration. Fortunately, there exists
a simpler way to obtain $ I_\text{R} $ which does not need
to consider the explicit value of the turning points $ r_{\min} $ and $ r_{\max} $. If
\be
I = \frac{2}{2\pi} \int_{r_{\min}}^{r_{\max}} dr \left( A + \frac{2B}{r} + \frac{C}{r^2} + \frac{D_1}{r^3} + \frac{D_2}{r^4} + \frac{D_3}{r^5} \right)^{\frac{1}{2}}
\ee
where $ A < 0, B > 0, C < 0, D_1 = \mc{O}(\epsilon), D_2 \text{ and } D_3 = \mathcal{O}(\epsilon^2) $, we have~\cite{damour1988higher}
\begin{equation}\label{eq::IRCal}
\begin{aligned}
I =& \frac{B}{\sqrt{-A}} - \sqrt{-C} \Bigg[ 1 - \frac{B}{2C^2} \left( D_1 - \frac{3B D_2}{2C}  + \frac{15 D_1^2 B}{8C^2}\right) -\\
& \frac{A}{4C^2} \left(D_2 - \frac{3 D_1^2}{4C}\right) + \frac{3B}{4C^3} \left( A - \frac{5B^2}{3C} \right) D_3 \Bigg] + \mc{O}(\epsilon^3).
\end{aligned}
\end{equation}
Using Eq.~\eqref{eq::IRCal}, we can calculate $I_{\rm{R}}$ in Eq.~\eqref{eq::IR}.

\subsection{Radial action variable of the EOB system at 3PM order}\label{Sec::2.2}

In this subsection, we evaluate the effective radial action $I^{\text{eff}}_\text{R}$ associated with a test body of rest mass $m_0$ moving in an effective curved spacetime with metric $g^\text{eff}_{\mu\nu}$. In the EOB framework \cite{damour1PM2016,damour1988higher}, one models the conservative dynamics of a nonspinning binary by mapping it to geodesic motion in an effective geometry. Accordingly, the spinless two-body system is represented by a test particle moving in an effective metric, which we take to be of the form
\be
ds^2_{\text{eff}} = -A(R)dt^2 + B(R)dR^2 + C(R)R^2\left(d\Theta^2 + \sin^2 \Theta d\Phi^2\right).
\ee
At the third order of $ G $, $ A(R) $ and $ B(R) $ can be expanded as
\begin{equation}\label{eq::AB}
\begin{aligned}
A(R) &= 1 + a_1 \frac{2GM_0}{R} + a_2 \left( \frac{GM_0}{R} \right)^2 + a_3 \left( \frac{GM_0}{R} \right)^3,\\
B(R) &= 1 + b_1 \frac{2GM_0}{R} + b_2 \left( \frac{GM_0}{R} \right)^2+b_3 \left( \frac{GM_0}{R} \right)^3,
\end{aligned}
\end{equation}
where $a_i, b_i (i=1,2,3) $ are dimensionless parameters and $ M_0 $ is a mass parameter which denotes the mass of the effective background. The function $ C(R) $ may be fixed either to $ C(R) = 1 $ in Schwarzschild coordinates or to $ C(R) = B(R) $ in isotropic coordinates.

\par For a test particle moving in the effective background spacetime, the effective Hamilton-Jacobi equation reads
\be\label{eq::HJEOB}
g^{\mu\nu}_{\text{eff}} \frac{\partial S_{\text{eff}}}{\partial x^\mu} \frac{\partial S_{\text{eff}}}{\partial x^\nu} + m_0^2 = 0.
\ee
In the equatorial plane ($ \Theta = \frac{\pi}{2} $), $ S_{\text{eff}} $ reduces to
\be\label{eq::Seff}
S_{\text{eff}} = -\mc{E}_0 t + L_0 \Phi + S^0_\text{R}(R, E, L),
\ee
where $ \mc{E}_0 $ and $ L_0 $ are the effective energy and the angular momentum, respectively.
By substituting Eq.~\eqref{eq::Seff} into \eqref{eq::HJEOB}, we obtain
\be
\frac{\mr{d} S^0_\text{R}}{\mr{d} R} = \sqrt{\mc{R}_0(R, \mc{E}_0, L_0)},
\ee
where
\be \label{eq::R0}
\mc{R}_0(R, \mc{E}_0, L_0) = \frac{B(R)}{A(R)} \mc{E}_0^2 - B(R) \left( m_0^2 + \frac{L_0^2}{C(R) R^2} \right).
\ee
Based on the definition of the effective radial action variable $I^{\text{eff}}_\text{R}$
\be\label{eq::IReff}
I^{\text{eff}}_\text{R}(\mc{E}_0, L_0) = \frac{2}{2\pi} \int_{R_{\min}}^{R_{\max}} \mr{d}R \sqrt{\mc{R}_0(R, \mc{E}_0, L_0)},
\ee
we calculate the effective radial action variable $I^{\text{eff}}_\text{R}$ in the isotropic gauge in the following.

\par For this case, $ C(R) = B(R) $. By substituting Eq.~\eqref{eq::AB} into \eqref{eq::R0},  we get
\be
\mc{R}_0(R, \mc{E}_0, L_0) = \frac{A_0 R^2 - L^2_0}{R^2} + A_1 \left( \frac{G}{R} \right) + A_2 \left( \frac{G}{R} \right)^2+ A_3 \left( \frac{G}{R} \right)^3+ A_4 \left( \frac{G}{R} \right)^4,
\ee
where
\begin{equation}
\begin{aligned}
A_0 &= \mc{E}^2_0 - m^2_0, \\
A_1 &= 2M_0 \left[ -m^2_0 b_1 + \mc{E}^2_0 \left( b_1-a_1 \right) \right], \\
A_2 &= M_0^2 \left[ -b_2 m^2_0 + \mc{E}^2_0 \left( 4a_1^2 - a_2 + b_2 - 4a_1b_1 \right) \right],\\
A_3 &= M_0^3 \left[ -b_3 m_0^2 - \mc{E}_0^2 \left( 8a_1^3 + a_3+  2a_1 \left(b_2-2a_2 \right)-b_3-8a_1^2b_1+2a_2b_1  \right) \right], \\
A_4 &= 0.
\end{aligned}
\end{equation}
Using Eq.~\eqref{eq::IRCal}, we can calculate $I_{\rm{R}}^{\rm{eff}}$ in Eq.~\eqref{eq::IReff}.

\subsection{Resolving the 3PM mapping via higher-order radial action}\label{Sec::2.3}
Using Eq.~\eqref{eq::IRCal}, we find that the coefficients of both $I_{\rm{R}}$ in Eq.~\eqref{eq::IR} and $I_{\rm{R}}^{\rm{eff}}$ in Eq.~\eqref{eq::IReff} vanish at order $\mathcal{O}(G^3)$. While this initially seems to suggest an inability of the radial action to fix the effective metric at the 3PM order, a more rigorous analysis reveals that the complete dynamical information of the potential at $\mathcal{O}(G^3)$ (denoted by $P_3$ or $A_3$) is properly encoded at $\mathcal{O}(G^4/L^3)$ in the expansion of $I_{\rm{R}}$. Specifically, since $D_2$ vanishes identically in both the real and EOB systems, the expansion of Eq.~\eqref{eq::IRCal} up to $\mathcal{O}(\epsilon^2)$ reduces to include only the first-order correction from $D_1$:
\begin{equation}
    I = \frac{B}{\sqrt{-A}} - \sqrt{-C} + \frac{B D_1}{2(-C)^{3/2}} + \mathcal{O}(\epsilon^2).
    \label{eq::IR_higher}
\end{equation}
Applying this to both the real two-body system and the EOB system, we can explicitly identify the mapping between their radial actions. Substituting the corresponding coefficients into Eq.~\eqref{eq::IR_higher} and performing an asymptotic expansion in the large angular momentum limit ($L \gg G$), we have:
\begin{equation} \label{eq::IR_real}
    I_{\rm{R}}^{\text{real}} = \frac{P_1 G}{2\sqrt{-P_0}} - L + \frac{P_2 G^2}{2 L} + \frac{P_1 P_3 G^4}{4 L^3} + \dots
\end{equation}
for the real system, and
\begin{equation}\label{eq::IR_eff}
    I_{\rm{R}}^{\text{eff}} = \frac{A_1 G}{2\sqrt{-A_0}} - L_0 + \frac{A_2 G^2}{2 L_0} + \frac{A_1 A_3 G^4}{4 L_0^3} + \dots
\end{equation}
for the EOB system.

\par At this point, it is crucial to distinguish the physical PM order of the Hamiltonian from the higher powers of \(G\) generated in the expansion of the radial action. In the physical 3PM approximation, the Hamiltonian, and hence the radial effective potential, is strictly truncated at \(\mathcal{O}(G^3/r^3)\); no independent \(\mathcal{O}(G^4/r^4)\) contribution is included. However, because the radial action \(I_{\rm R}\) is a nonlinear functional of the radial effective potential, its asymptotic expansion naturally generates cross-terms. The \(\mathcal{O}(G^4/L^3)\) terms in Eqs.~\eqref{eq::IR_real} and \eqref{eq::IR_eff} are nonlinear products of coefficients already determined through 3PM, namely \(P_1P_3\) and \(A_1A_3\). Therefore, matching the radial actions at this order does not introduce 4PM dynamics or exceed the physical 3PM accuracy; rather, it extracts the complete 3PM information encoded in the nonlinear expansion of the radial action.
\par By enforcing the correspondence principle $I_{\rm{R}}^{\text{real}} = I_{\rm{R}}^{\text{eff}}$ with $L=L_0$, we can match the terms order by order. While the $\mathcal{O}(G^3)$ terms on both sides are indeed zero, the $\mathcal{O}(G^1)$, $\mathcal{O}(G^2)$ and $\mathcal{O}(G^4)$ terms provide constraints:
\begin{equation}
\begin{aligned}
    \frac{P_1 G}{2\sqrt{-P_0}} &= \frac{A_1 G}{2\sqrt{-A_0}}, \\
    \frac{P_2 G^2}{2 L} &= \frac{A_2 G^2}{2 L}, \\
    \frac{P_1 P_3 G^4}{4 L^3} &= \frac{A_1 A_3 G^4}{4 L^3}.
\end{aligned}
\end{equation}
We finally derive the correspondence:
\begin{equation}\label{eq::RadialMatch}
\begin{aligned}
    A_1 =& \frac{P_1\sqrt{A_0}}{\sqrt{P_0}}, \\
    A_2 &= P_2, \\
    A_3 \sqrt{A_0} &= P_3 \sqrt{P_0}.
\end{aligned}
\end{equation}
Remarkably, this result is identical to the expression obtained through the precession angle matching in Sec.~\ref{Sec::3.1}. This demonstrates that the radial action variable method is consistent with the precession angle method in determining the EOB metric.

\section{Effective metric based on precession angle}\label{sec::PrecessionAngle}

\par Building on earlier work, Damour established the energy mapping between a real two-body system and the effective-one-body (EOB) description at the 1PM and 2PM levels \cite{damour1PM2016,damour2PM2019}. Later, Antonell, Jing et al. extended Damour's result to 3PM and 4PM level \cite{antonelli2019energetics,Jing2023okh}. It should be mentioned that all these works are based on the calculation of the scattering deflections.  In contrast, our analysis targets bound states, for which the relevant observable is the precession angle rather than the scattering angle. To independently verify the subtle constraints derived from the higher-order radial action in Sec.~\ref{Sec::2.3}, we now turn to a direct physical observable of bound states: the precession angle. In this section, we extract the effective metric at 3PM by analyzing the precession angles of closed orbits and demonstrate that it yields exactly the same mapping.

\subsection{Precession angle of real two-body system at 3PM order}\label{Sec::3.1}

\par Using the Hamilton-Jacobi formalism, it can be shown (e.g., see Ref.~\cite{bern2019long}) that the total change in the angle coordinate $\phi$ for a scattering orbit is
\be
\chi = -\pi + 2J \int_{r_{\min}}^{\infty} \frac{dr}{r^2 \sqrt{p_r(r)^2}},
\ee
where the minimum distance between the two particles by $ r_{\min} $, at which point $p_r(r_{\min})=0$. Another form of the scattering angle can be seen in Ref.~\cite{damour2PM2019}.

\par For a bound state the scattering angle should be replaced by the precession angle which is given by
\be\label{eq::Angle}
\psi = -2\pi + 2J \int_{r_1}^{r_2} \frac{dr}{r^2 \sqrt{p_r(r)^2}}.
\ee
Here $ r_1 $ and $ r_2 $ are the roots of $p_r(r)^2 = 0$. In a bound state the radial momentum $p$ reverses sign at two turning points---the periastron and apastron. Given the form of $p_r^2$ in Eq.~\eqref{eq::prEp}, the computation reduces to evaluating the integral
\be\label{eq::AngleCal}
I = \int_{u_c}^{u_b} \frac{du}{\sqrt{(u_a - u)(u_b - u)(u - u_c)}}
\ee
at $G^3$ order. Here we assume the roots are ordered as $ u_a > u_b > u_c $. The integral for $u$ is performed on the interval $u_c < u < u_b$. The solution at $G^3$ order is
\be
I = \frac{1}{\sqrt{u_a - u_c}} K(k^2),
\ee
where $K(k^2)$ represents the complete elliptic integral of the first kind
\be
k^2 = \frac{u_b - u_c}{u_a - u_c}, \quad
K(k^2) = \int_0^{\pi/2} (1 - k^2 \sin^2 \phi)^{-1/2} \mr{d}\phi.
\ee
For the case where $k^2$ is small $(k^2 \ll 1)$
\be
K(k^2) = \frac{\pi}{2} \left( 1 + \frac{k^2}{4} \right) + \mathcal{O}(k^4)
\ee
Using Eq.~\eqref{eq::AngleCal}, the precession angle Eq.~\eqref{eq::Angle} may be calculated as
\be\label{eq::AngleReal}
\psi=P_2 \pi \left(\frac{G}{L} \right)^2 + P_3\sqrt{P_0}  \pi \left(\frac{G}{L}\right)^3 + \mathcal{O}\left(G^4\right)
\ee
with $P_0, P_2, P_3$  given in Eqs.~\eqref{eq::PValue}. At $\mathcal{O}(G^2)$ order, the result agrees with Ref.~\cite{he2021energy}.

\subsection{Precession angle of EOB system at the 3PM order}\label{Sec::3.2}

We now compute the effective precession angle $ \psi_0 $ in isotropic gauge $ (C(R) = B(R)) $. For an EOB system, we have obtained, in Sec.\ref{Sec::2.2}, that
\be
P^2_\text{R} = \frac{\mr{d}S^0_\text{R}}{\mr{d}R} = \frac{B(R)}{ A(R)} \mc{E}^2_0 - B(R) \left( m^2_0 + \frac{L^2_0}{C(R) R^2} \right),
\ee
where $ A(R), B(R) $ are given by Eq.~\eqref{eq::AB} with expansion coefficients.

\par Within isotropic coordinates, direct calculation shows that $ P^2_\text{R} $ at the 3PM order is
\be
P^2_\text{R} = \frac{A_0 R^2 - J^2_0}{R^2} + A_1 \left( \frac{G}{R} \right) + A_2 \left( \frac{G}{R} \right)^2+ A_3 \left( \frac{G}{R} \right)^3,
\ee
where
\begin{equation}
\begin{aligned}
A_0 &= \mc{E}^2_0 - m^2_0, \\
A_1 &= 2M_0 \left[ -m^2_0 b_1 + \mc{E}^2_0 \left( b_1-a_1 \right) \right], \\
A_2 &= M_0^2 \left[ -b_2 m^2_0 + \mc{E}^2_0 \left( 4a_1^2 - a_2 + b_2 - 4a_1b_1 \right) \right],\\
A_3 &= M_0^3 \left[ -b_3 m_0^2 - \mc{E}_0^2 \left( 8a_1^3 + a_3+  2a_1 \left(b_2-2a_2 \right)-b_3-8a_1^2b_1+2a_2b_1  \right) \right].
\end{aligned}
\end{equation}
Here we emphasize that it is in this isotropic gauge, rather than in the Schwarzschild areal-radius gauge \(C(R)=1\), that we can obtain the above form, down to only $R^{-3}$, which greatly simplifies the analytical calculation of the integral. From Eq.~\eqref{eq::AngleCal}, the effective precession angle can be computed and the result is
\begin{align}\label{eq::AngleEff}
\psi^{\rm{eff}} =& A_2 \pi \left(\frac{G}{L}\right)^2 + A_3\sqrt{A_0}  \pi \left(\frac{G}{L}\right)^3 + \mathcal{O}(G^4)
\nnm\\
\quad =& \pi\left(\frac{G}{L}\right)^2 M_0^2  \left[ -b_2 m_0^2 + \mc{E}_0^2 \left(4a_1^2 - a_2 + b_2 - 4 a_1b_1\right) \right] +
\nnm\\
& \pi\left(\frac{G}{L}\right)^3 \sqrt{\mc{E}_0^2 - m_0^2}  M_0^3 \left[ -b_3 m_0^2 - \mc{E}_0^2 \left( 8a_1^3 + a_3+ 2a_1 \left( b_2-2a_2 \right)-b_3-8a_1^2b_1+2a_2b_1  \right) \right].
\end{align}
By using the correspondence principle, we identify the effective precession angle $\psi^{\rm{eff}}$ in Eq.~\eqref{eq::AngleEff} with the real two-body one $\psi$ in Eq.~\eqref{eq::AngleReal}. Focusing on the $\mathcal{O}(G^2)$ and $\mathcal{O}(G^3)$ terms, it is straightforward to find that this matching condition yields the same correspondence relations as those derived from the radial action in Eq.~\eqref{eq::RadialMatch}. This confirmation reinforces the consistency between the radial-action and precession-angle correspondences. These matching relations constrain the effective metric coefficients, while their unique specification additionally requires a choice of metric parametrization.

\subsection{Effective metric at 3PM order based on precession angle}
Now we construct the effective metric by applying the correspondence in Eq.~\eqref{eq::RadialMatch} together with the energy map Eq.~\eqref{eq::EnergMap} between effective energy $\mc{E}_0$ and $\mc{E}$. Implementing these maps does not fully fix the effective metric, since at each PM order $i$ a single equation must accommodate two parameters $(a_i, d_i)$. A typical convention is to fix $a_1=-1$; with this in place, one then isolates the $G^2$ order contributions and compares coefficients of the same powers of $\mc{E}$. This determines $a_2$ and $b_2$ \cite{he2021energy}.

However, this approach encounters some difficulties in dealing with the coefficients $a_3, b_3$. Therefore, within the isotropic gauge \(C(R)=B(R)\), we adopt a Schwarzschild-like parametrization to fix the remaining freedom by choosing \(a_1=-1\) and \(b_i=0~(i=2,3)\). Moreover, following Ref.~\cite{damour1PM2016}, we choose
\begin{subequations}\label{Eq::coefficientsb}
\begin{align}
b_1 &= 1,\\
m_0 &= \frac{m_1m_2}{m_1+m_2},\\
M_0 &= m_1+m_2,\\
\mc{E}_0 &= \frac{\mc{E}^2-m_1^2-m_2^2}{2(m_1+m_2)}.
\end{align}
\end{subequations}
After determining the above coefficients, we can determine the remaining $a_i$ coefficients
\begin{subequations}\label{Eq::coefficientsb}
\begin{align}
    a_2 &= 8-\frac{P_2}{\mc{E}_0^2M_0^2}
,\\
    a_3 &= 16-6 \left( 8-\frac{P_2}{\mc{E}_0^2M_0^2} \right)-\frac{\sqrt{P_0}P_3}{\mc{E}_0^2\sqrt{\mc{E}_0^2-m_0^2}M_0^3}.
\end{align}
\end{subequations}
\par After completing the calculation of the above coefficients, we have completed the conversion of the real two-body problem to the effective single-body problem. The result is represented by an effective metric, and the coefficients in the metric are expressed by the physical parameters of the real two-body system.

\section{Comparison with existing results}
In the isotropic gauge with a Schwarzschild-like parametrization, we have shown that the effective metric is
\begin{align}
    ds^2_{\text{eff}} &= -A(R)dt^2 + B(R)dR^2 + C(R)R^2 \left( d\Theta^2 + \sin^2 \Theta d\Phi^2 \right), \\
    A(R) &= 1 - \frac{2GM_0}{R}+a_2 \left( \frac{GM_0}{R} \right)^2+a_3 \left( \frac{GM_0}{R} \right) ^3,\\
    B(R) &=C(R) = 1 + \frac{2GM_0}{R}.
\end{align}
Defining the reduced (dimensionless) quantities for later convenience
\begin{gather*}\label{reducedHu}
  \hat{\mc{E}}_0=\frac{\mc{E}_0}{m_0},\quad \hat{\mc{E}}_\mr S=\frac{\mc{E}_\mr S}{m_0},\quad u=\frac{GM_0}{r},\quad \nu=\frac{m_0}{M_0},\\
  ~\hat{p}_r = \frac{p_r}{m_0},~ l \equiv \hat{p}_\phi = \frac{L}{GM_0m_0},~\Gamma=\sqrt{1+2\nu(\hat{\mc{E}}_0-1)}.
\end{gather*}
Then the Schwarzschild-geodesic Hamiltonian (for a test particle of mass $m_0$) is given in isotropic coordinates in the equatorial plane by
\be \label{Eq::HatEs}
\hat{\mc{E}}_\mr{S}^2 = (1-2u) \left[1 + (1-2u)l^2 u^2 + (1-2u) \hat{p}_r^2 \right].
\ee
Moreover, the squared reduced effective Hamiltonian at the 3PM order reads
\be \label{Eq::HatEeff}
\hat{\mc{E}}^2_{0} = (1 - 2u+a_2u^2+a_3u^3) \left[ 1 + \frac{1}{1 + 2u}(\hat{p}_r^2+l^2u^2) \right] .
\ee
Eq.~\eqref{Eq::coefficientsb} then turns into:
\begin{align}\label{Eq::a2_Sch}
    a_2 &=8+\frac{3-15\hat{\mc{E}}_0^2}{2\hat{\mc{E}}_0^2\Gamma},\\
    \label{Eq::a3_Sch}
    a_3 &= 16 - 6 a_2 - \frac{1}{6 \hat{\mc{E}}_0^2 \sqrt{-1 + \hat{\mc{E}}_0^2}}\sqrt{\frac{-1 + \hat{\mc{E}}_0^2}{\Gamma^2}}\Bigg[ \frac{9 (-1 + 2 \hat{\mc{E}}_0^2) (-1 + 5 \hat{\mc{E}}_0^2) (-1 + \Gamma)}{-1 + \hat{\mc{E}}_0^2} + \nnm\\
    & 3 (-1 + 18 \hat{\mc{E}}_0^2) \Gamma - \frac{8 \hat{\mc{E}}_0 (25 + 14 \hat{\mc{E}}_0^2) \nu}{\Gamma} - \frac{48 (-3 - 12 \hat{\mc{E}}_0^2 + 4 \hat{\mc{E}}_0^4) \nu \sinh^{-1}\left( \sqrt{\frac{-1 + \hat{\mc{E}}_0}{2}} \right)}{\sqrt{-1 + \hat{\mc{E}}_0^2} \Gamma} \Bigg].
\end{align}
By evaluating Eq. \eqref{Eq::HatEeff} at the 1PM order and substituting the result into Eq. \eqref{Eq::a2_Sch}, $a_2$ is determined as a function of $u$, $\hat{p}_r$, and $l$. Subsequently, substituting this expression back into Eq. \eqref{Eq::HatEeff} yields $\hat{\mathcal{E}}_0$ to the 2PM order. Inserting the resulting 2PM energy into Eq. \eqref{Eq::a3_Sch} allows us to obtain $a_3(u, \hat{p}_r, l)$.

\par Once we get $a_2$ and $a_3$ in terms of $u$, $\hat{p}_r$, and $l$ we are now in a position to compare our EOB predictions for the binding energy of a two-body system on a quasi-circular orbit with results from existing PM-EOB models. To this end, we introduce $e=\frac{\mc{E}-M_0}{m_0}$, where $\mc{E}$ is the total energy of a two-body system, and use adiabatic approximation to investigate the energy-angular momentum ($e-l$) relation.
\par
\begin{figure*}[htbp]
\begin{tabular}{cc}
  \includegraphics[width=0.45\textwidth]{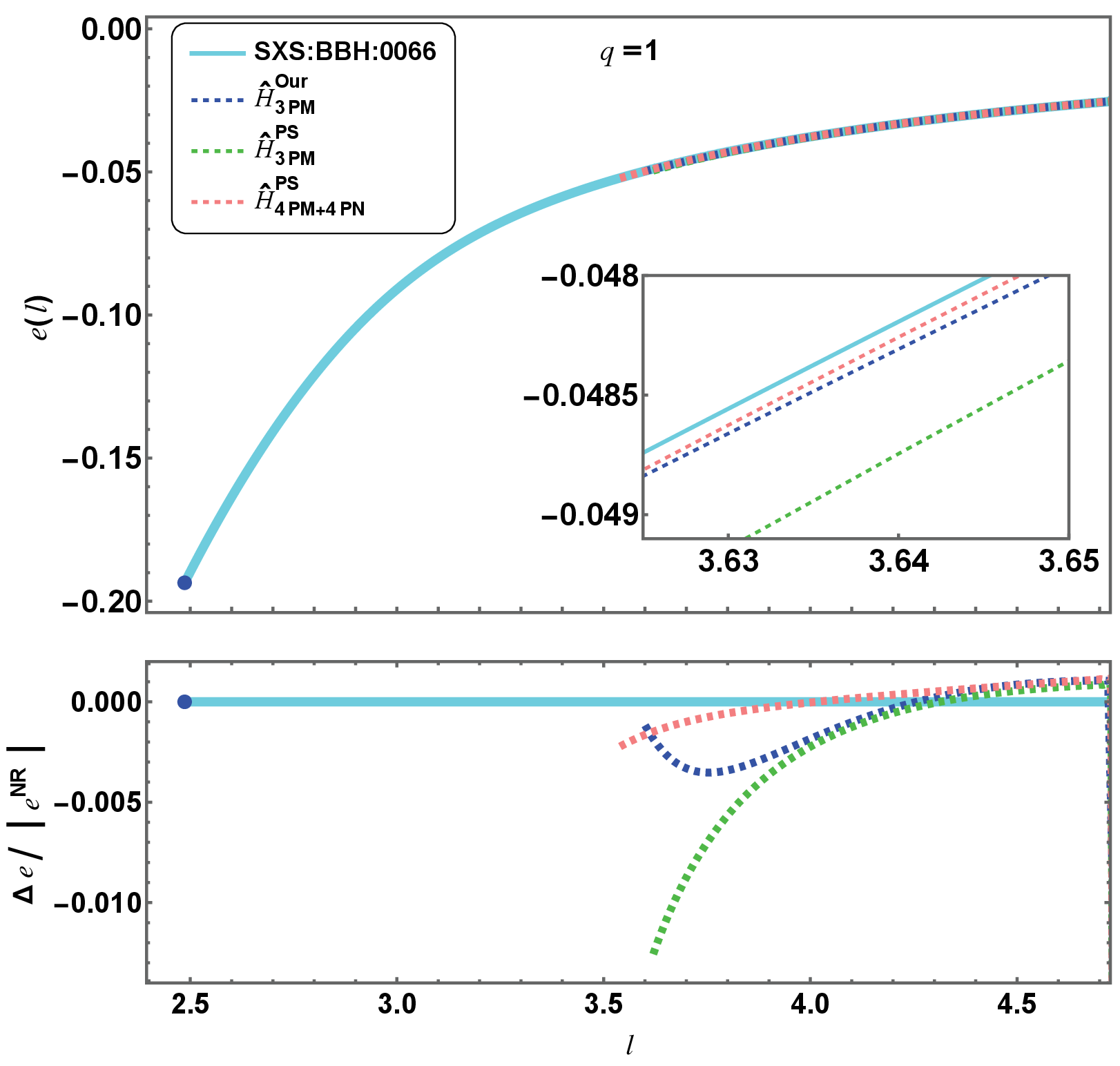}&
  \includegraphics[width=0.45\textwidth]{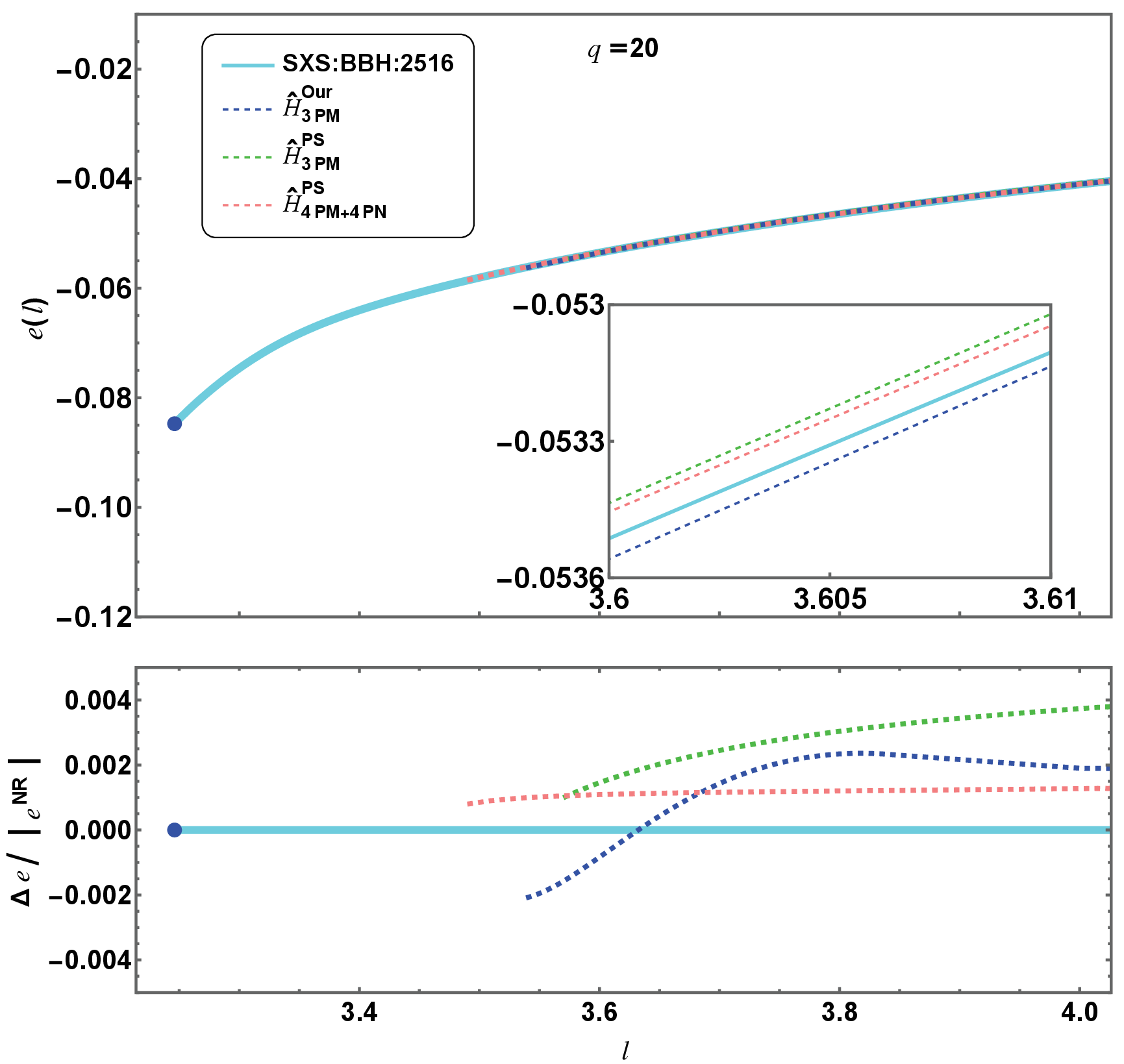}
\end{tabular}
    \caption{{\bf Energetics of PM Hamiltonians.} We compare to NR the binding energy $e$ as a function of dimensionless angular momentum $l=L/(Gm_0M_0)$ from PM-EOB Hamiltonians for a nonspinning binary black hole with mass ratio $q=1$ (left panel) and $q=20$ (right panel). The ends of the curves mark the ISCOs, when present in the corresponding two-body dynamics. The NR binding energy is in cyan. In the lower panel we show the fractional difference between the approximants and the NR result. }
    \label{Fig::1}
\end{figure*}

 In Fig.~\ref{Fig::1}, we compare the $e\text{-}l$ relation of our proposed 3PM EOB Hamiltonian against the previous 3PM formulation, using the recent 4PM EOB Hamiltonian and Numerical Relativity (NR) binding energy as benchmarks. We observe that the new Hamiltonian yields results that are more consistent with these high-precision baselines, notably achieved without introducing Finsler-type or PN corrections. As the binary evolves from inspiral toward merger, corresponding to a decrease in angular momentum $l$, the deviation between the different approximations becomes increasingly distinct.

\section{Summary and discussion}\label{sec::Summary}

\par The effective-one-body (EOB) framework provides a powerful method for modeling gravitational waves from binary black hole systems. When calibrated to numerical relativity, the resulting EOBNR models have become central to gravitational-wave data analysis. Traditionally, EOB model is built from post-Newtonian (PN) information about the relativistic two-body dynamics. In parallel, the post-Minkowskian (PM) expansion offers a complementary approximation scheme for the same problem.

\par A PM-based formulation of EOB has recently been developed. At first post-Minkowskian (1PM) order, by matching the scattering angles of the real two-body system and its effective counterpart, one obtains the energy mapping between the real and effective descriptions, as expressed in Eq.~\eqref{eq::EnergMap}. In 2019, Damour derived the 2PM Hamiltonian by computing the effective scattering angle using a generalized mass-shell condition that includes a Finsler-type term \cite{damour2PM2019}. Subsequently, Antonelli and collaborators extended this program to third post-Minkowskian (3PM) order and analyzed the energetics of two-body Hamiltonians at that accuracy \cite{antonelli2019energetics}. To establish a correspondence between gravitational observables in scattering processes and adiabatic invariants in bound orbits, K{\"a}lin and Porto introduced a conceptual framework \cite{scattertobound}.

\par The conventional route to study bound states relies on matching scattering angles and then applying the Ka\"lin-Porto map. Here, we pursue a different strategy to analyze bound states and to construct the effective metric and Hamiltonian within the EOB framework at third post-Minkowskian (3PM) order. We first evaluate the radial action for bound orbits and establish the 3PM correspondence through it. Consistently, we also extract the precession angle from the real two-body system and construct the effective metric in the EOB formalism at 3PM accuracy based on it. When constructing the effective metric, we work in the isotropic gauge and adopt a Schwarzschild-like parametrization to fix the remaining freedom. This contrasts with Damour's approach, which introduces a Finsler-type contribution. Notably, our 3PM effective metric depends explicitly on the relativistic energy \(E\) of the real binary. This feature is physically well motivated: different values of \(E\) correspond to distinct initial configurations, which can lead to different gravitational-wave emissions.  Here, the energy dependence should not be interpreted as defining a universal spacetime geometry fixed solely by the intrinsic binary parameters. Rather, for each fixed value of the conserved energy \(E\), the metric provides a state-dependent effective representation of the conservative two-body dynamics, while changing \(E\) selects a different member of this family of effective metrics. Since \(E\) is conserved, the corresponding effective metric remains fixed along a given conservative bound orbit. In this sense, the expression ``effective metric for bound-state dynamics'' refers to an auxiliary, energy-shell-dependent geometrization of the two-body dynamics rather than to a unique physical spacetime geometry of the binary.  As a validation, our effective Hamiltonian agrees with existing results at 2PM order, confirming consistency at that accuracy.

\par In this paper, we have extended Damour's post-Minkowskian (PM) effective-one-body program from scattering configurations to bound states. Our analysis proceeds directly from bound-state dynamics and yields an explicit effective metric, providing a practical tool for modeling gravitational-wave emission from compact binaries. As future directions, we plan to construct the energy map and refine the effective metric at higher PM orders, and to generalize the present framework to binaries with spin.

\acknowledgments
This work is supported by the National Key Research and Development Program of China (Nos. 2021YFC2203001), the National Natural Science Foundation of China (Nos.~12475049, 12547143 and 12275350) and the Natural Science Foundation of Hunan Province (No.2023JJ30179).

\bibliography{refs}

\end{document}